# WAVELET BASED AUTHENTICATION/SECRET TRANSMISSION THROUGH IMAGE RESIZING (WASTIR)

Madhumita Sengupta[1] and J. K. Mandal[2]

[1, 2] Department of Computer Science and Engineering, University of Kalyani
Kalyani, Nadia, Pin. 741235, West Bengal, India
[1] madhumita.sngpt@gmail.com  [2] jkm.cse@gmail.com

## ABSTRACT

*The paper is aimed for a wavelet based steganographic/watermarking technique in frequency domain termed as WASTIR for secret message/image transmission or image authentication. Number system conversion of the secret image by changing radix form decimal to quaternary is the pre-processing of the technique. Cover image scaling through inverse discrete wavelet transformation with false Horizontal and vertical coefficients are embedded with quaternary digits through hash function and a secret key. Experimental results are computed and compared with the existing steganographic techniques like WTSIC, Yuancheng Li's Method and Region-Based in terms of Mean Square Error (MSE), Peak Signal to Noise Ratio (PSNR) and Image Fidelity (IF) which show better performances in WASTIR.*

## 1. INTRODUCTION

Digital steganography plays a vital role in verifying the integrity of digital information. Watermarking can achieve copyright protection, ownership verification or secret message transmission depending on the ways of data hiding. Various techniques are available categorized under spatial domain [2, 5, 13], frequency domain [3, 5] or composite domain [4]. Like LSB, PDV, SVD [1], DCT [4], DFT, DWT [1, 3, 5, 12], techniques of arbitrary resizing [12] and many more.

Necessity of security is a major concern of today's researchers, despite of various works in this area still there is dearth of investigation in the subject. There are two major improvements seen in the proposed one. Firstly the degradation of the quality of the cover image is less compared to other techniques. Secondly the payload has been increased considerably without degrading the fidelity of the image.

Various parametric tests have been performed and results obtained are compared with most recent existing techniques such as, WTSIC [3], Yuancheng Li's Method [12] and Region-Based watermarking [13], based on Mean Square Error (MSE), Peak Signal to Noise Ratio (PSNR) and Image Fidelity (IF) analysis [6] which shows a consistent relationship with the existing techniques.

Section 2 of the paper deals with the proposed scheme. Resizing through discrete wavelet transformation technique with embedding has been described in section 3, Results and discussions are outlined in section 4, conclusions are drawn in section 5 and that of references are given at end.

## 2. THE SCHEME

Digital images are represented as a two dimensional matrix of pixel intensity values in the form of integer representation in the range of 0 to 255. The proposed WASTIR technique is divided into two major groups of activity shown in figure 1 and 2.

The schematic representation of activity at sender is shown in figure 1. The cover image of dimension N x N passes through the process of inverse DWT to generate translated image with predefined horizontal, vertical and diagonal information of N x N dimension each, as describe in section 3.3. The secret message/image converts its pixel intensity values from decimal to quaternary with radix 4, as shown in figure 4.

On applying 2 x 2 masks over translated image $R_{00}$, $R_{01}$, $R_{10}$ and $R_{11}$ values are fetched to embed two digits (4 bits) of quaternary system based on hash function as explained in section 3.4. Fidelity adjustment has been incorporated to preserve the visual quality of cover image which is discussed in section 3.5 to generate stego-image.

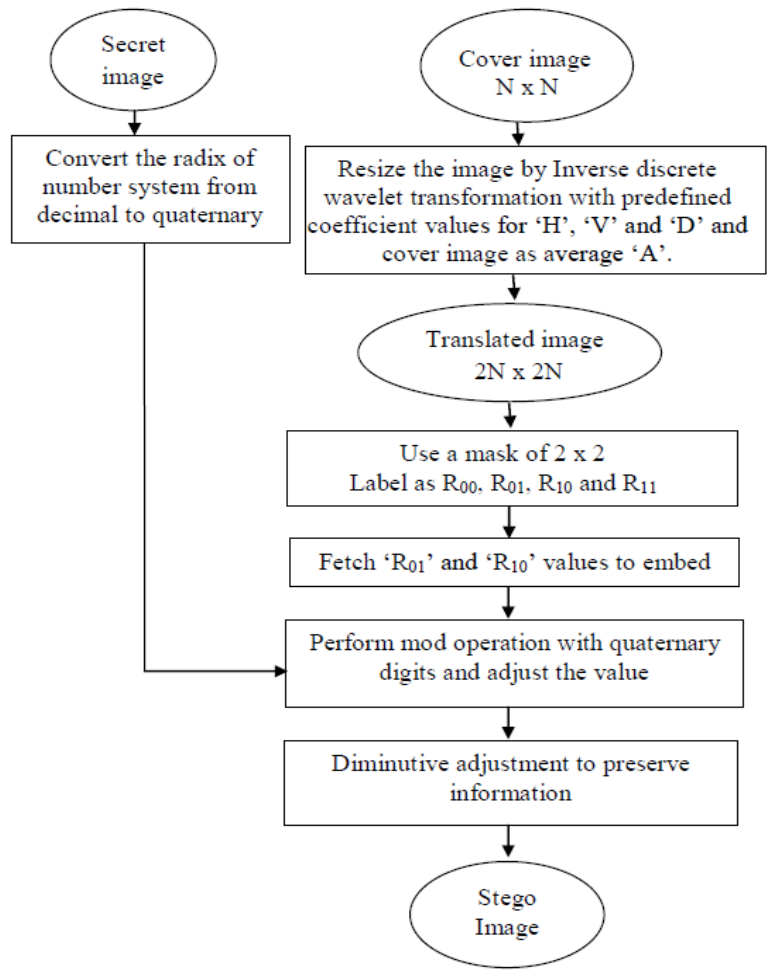

Figure 1.   Embedding Algorithm

During decoding the stego-image passes through the extraction algorithm based on same hash function to regenerate the secret message/image. On authentication or secret message generation the stego-image passes through the forward DWT as shown in section 3.1. The output of forward DWT generates the average coefficients of N x N dimension that is exactly the cover image taken by sender with MSE zero and PSNR infinity. Embedding process is shown in figure 2.

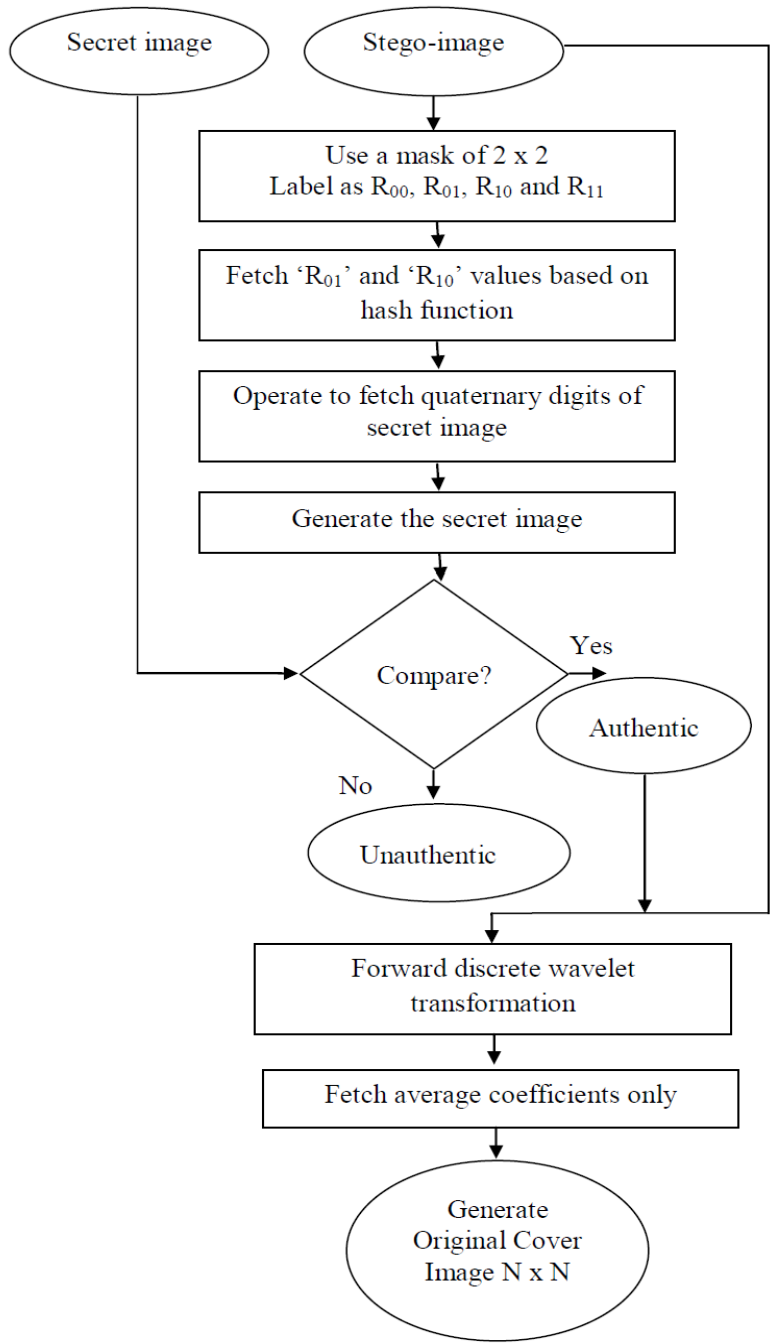

Figure 2.   Extraction/Authentication Algorithm

## 3. RESIZING THROUGH TRANSFORMATION AND EMBEDDING

In image processing each transform equation is available as pair which is reversible and termed as forward and inverse transformation respectively[7]. In Wavelet based forward transformation the image converts from spatial domain to frequency domain using (1) and (2), and in inverse transformation the reverse procedure is followed (3). Mathematically the image matrixes multiply with scaling function coefficients and wavelet function coefficients in the formation of transformation matrix [8].

$$Y_{Low}[k] = \sum_n x[n].h[2k-n] \quad (1)$$

$$Y_{High}[k] = \sum_n x[n].g[2k-n] \quad (2)$$

$$x[n] = \sum_{k=-\infty}^{\infty}(Y_{High}[k].g[2k-n]) + (Y_{Low}[k].h[2k-n]) \quad (3)$$

Where x[n] is original signal, h[x] is half band low pass filter, g[x] is Half band high pass filter, YLow[k] is output of high pass filter after sub sampling by 2, YHigh[k] is output of low pass filter after sub sampling by 2.

### 3.1 Forward Transformation

In the proposed technique Mallat based two-dimensional wavelet transform is used in order to obtain set of bi-orthogonal subclasses of images [9]. In two-dimensional wavelet transformation, a scaling function φ(x, y) is used represent by (4).

$$\varphi(x, y) = \varphi(x)\varphi(y) \quad (4)$$

and if ψ(x) is a one-dimensional wavelet function associated with the one–dimensional scaling function φ(x), three two dimensional wavelets may be defined as given in (5). Figure 3 represents functions in visual form.

$$\begin{aligned}\psi H\ (x,y) &= \varphi(x)\psi(y) \\ \psi V\ (x,y) &= \psi(x)\varphi(y) \\ \psi D\ (x,y) &= \psi(x)\psi(y)\end{aligned} \quad (5)$$

| Low resolution sub-image average 'A' $\psi(x, y)= \varphi(x)\varphi(y)$ | Horizontal Orientation sub-image 'H' $\psi^H(x,y)= \varphi(x)\psi(y)$ |
|---|---|
| Vertical Orientation sub-image 'V' $\psi^V(x, y)= \psi(x)\varphi(y)$ | Diagonal Orientation sub-image 'D' $\psi^D(x,y)=\psi(x)\psi(y)$ |

Figure 3.  Image decomposition in Wavelet transforms

Signal & Image Processing : An International Journal (SIPIJ) Vol.4, No.2, April 2013

As per Haar forward transform scaling function coefficients and wavelet function coefficients [8] H0 = ½, H1 = ½, G0 = ½ G1 = -½ are taken. In WASTIR stego-image at receiver side of dimension 2N x 2N on forward DWT breaks into four sub images as shown in figure 3. Each sub-image is of dimension N x N. Out of which low resolution/average sub image on comparison with is the original cover image taken by sender, generates MSE zero and that of PSNR infinity. Thus without hampering the original image, secret message/image can be transmitted through unsecured network, securely.

### 3.2 Inverse Transformation

Inverse transformation is just the reverse of the forward transformation with column transformation done first followed by row transformation. But the coefficient values are different for column/row transformation matrices. The coefficient for reverse transformation are $H_0 = 1$, $H_1 = 1$, $G_0 = 1$, $G_1 = -1$ [8]. In the first phases of proposed WASTIR technique, original image of dimension N x N passes through inverse transform to generate translated image matrix of dimension 2N x 2N.

For any inverse wavelet transformation (equation (3) of section III) four equal sized coefficient matrix is required such as low resolution sub-image 'A', horizontal orientation sub-image 'H', vertical orientation sub-image 'V' and diagonal orientation sub-image 'D' as shown in figure 3, they are output of forward wavelet transformation. Thus to perform inverse wavelet transformation without forward wavelet transformation three false coefficient matrix 'H', 'V', and 'D' of same size is required with original image as 'A' for the process of resizing or scaling.

### 3.3 Resizing/Scaling

To scale image by a factor T, up-sampling by a factor of T is performed. This up-sampling is done through discrete wavelet transformation, where four types of coefficient sets are voted through (3) to generate image of dimension TN x TN. The original cover image of dimension N x N is taken as average 'A', where as horizontal, vertical and diagonal coefficients are taken as predefined values. The space overhead will depend on the value of T.

Let us take an example where resize of the image has been done by a factor of 2, for which cover image of dimension 2 x 2 is taken and substitute with 2 x 2 dimension of horizontal, vertical and diagonal sub components each with a false value β. Then after inverse wavelet transformation, translated image generates with a dimension of 4 x 4 as shown in figure 4.

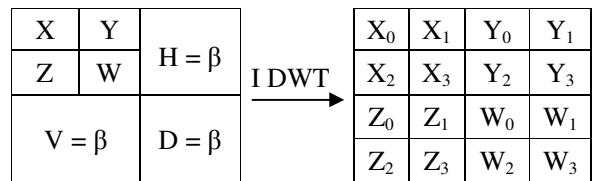

Figure 4.  Resizing of 2 x 2 cover image into 4 x 4 translated image through inverse wavelet transformation

Where β is a user define integer taken as a false horizontal, vertical and diagonal coefficients to perform IDWT.

### 3.4 Embedding

The secret image of dimension 512 x 256 is selected as shown in figure 8.k. This secret image is of 1048576 bits in size. Each pixel intensity value represented in the form of decimal number system in range of 0 to 255 has been converted to quaternary number system. The minimum

63

intensity value 0 in decimal can be represented as $(0000)_4$ in quaternary number system, and 255 the maximum intensity supported by our system of image can be represented as $(3333)_4$. Few examples with conversion technique are given in figure 5. After conversion of radix the intensity values of secret image is stored in a long chain of digits array ranging from 0 to 3.

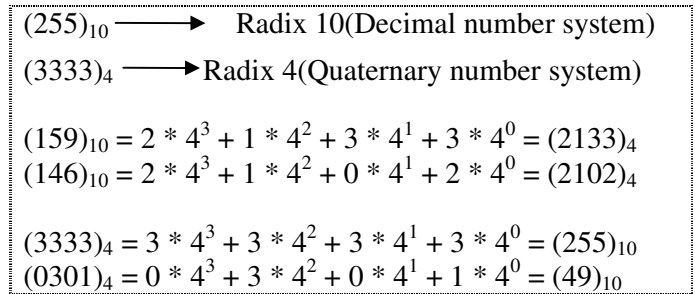

$(255)_{10} \longrightarrow$ Radix 10(Decimal number system)

$(3333)_4 \longrightarrow$ Radix 4(Quaternary number system)

$(159)_{10} = 2 * 4^3 + 1 * 4^2 + 3 * 4^1 + 3 * 4^0 = (2133)_4$
$(146)_{10} = 2 * 4^3 + 1 * 4^2 + 0 * 4^1 + 2 * 4^0 = (2102)_4$

$(3333)_4 = 3 * 4^3 + 3 * 4^2 + 3 * 4^1 + 3 * 4^0 = (255)_{10}$
$(0301)_4 = 0 * 4^3 + 3 * 4^2 + 0 * 4^1 + 1 * 4^0 = (49)_{10}$

Figure 5.  Example of conversion of number system

A 2 x 2 mask on translated image is used which provides four values at every pair of coordinates, labeled as $R_{00}$, $R_{01}$, $R_{10}$, and $R_{11}$. Out of four two diagonal coordinate pairs, $R_{01}$ and $R_{10}$ are selected for embedding.

The hash function with mod operation works together to select the embedding position and the embedded value, as per (6). This procedure is illustrated in figure 6. The hash function selects $R_{01}$ as the first embedding position. Then '$R_{01}$ mod 4' operation is performed and the difference is calculated between the secret quaternary digit and mod result. If the difference is positive and more than zero then the new value for the $R_{01}$ position is recalculated by subtracting the difference. Else if the difference is less than zero then the new value for the position is recalculated by adding the difference.

diff ='Secret digit' – ('old value of embedding position' % 4)
If (diff > 0 ) then
    'New value of embedding position'= 'Old value of
     embedding position'– (4 – diff)
if (diff < 0 ) then
    'New value of embedding position'= 'Old value of
    embedding position'+ (4 + diff)  (6)

Hence two secret digits 3 and 2 taken from the chain of secret digits are embedded through hash function into $R_{01}$ and $R_{10}$ based on generalized (6).

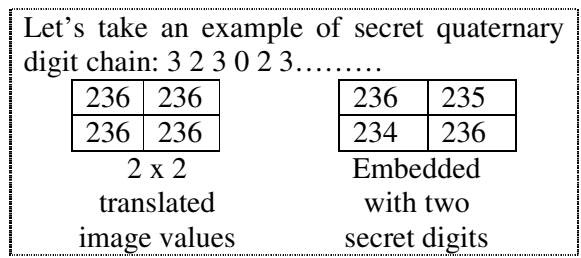

Let's take an example of secret quaternary digit chain: 3 2 3 0 2 3………

| 236 | 236 |     | 236 | 235 |
|-----|-----|-----|-----|-----|
| 236 | 236 |     | 234 | 236 |

2 x 2 translated image values     Embedded with two secret digits

if [('secret digit'– ($R_{01}$ % 4)) > 0] then
  [$R_{01}$= $R_{01}$ – (4 – (secret digit'– ($R_{01}$ % 4))
else if [('secret digit'– ($R_{01}$ % 4)) < 0] then
  [$R_{01}$= $R_{01}$ + (4 + (secret digit'– ($R_{01}$ % 4))

Figure 6.  Embedding two secret digits in 2 x 2 mask of translated image.

### 3.5 Adjustment

Diminutive adjustment is required to preserve the near original image intensity. The adjustment enforces equal amount of deduction from the diagonal value $R_{11}$. The embedded image peruses forward DWT without adjustment as shown in figure 7.c where the low resolution or average value changes from 236 to 235.25 and with adjustment as shown in figure 7.d, where the low resolution or average value retains at 236.

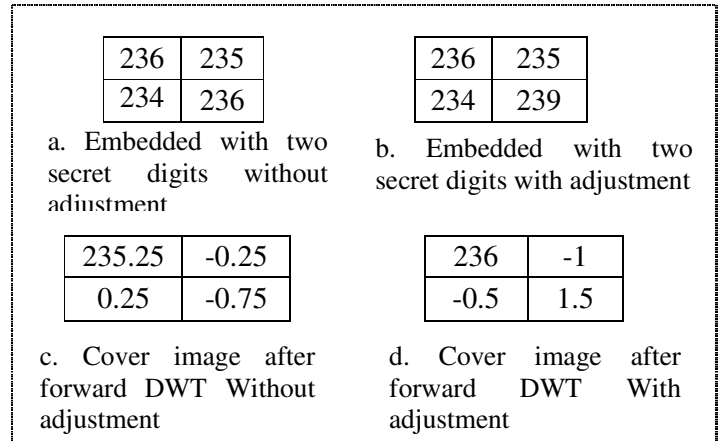

Figure 7.  Procedure of diminutive adjustment

From figure 7 it is clear that, to regenerate the original cover image the low resolution sub-image must be intact on forward DWT, after embedding also, which generates the requirement of adjustment.

Adjustment is a simple calculation of equating the total average before embedding and after embedding by readjusting the value of R11. In the above example of figure 6 and figure 7, the average of 2 x 2 translated image is 236. After embedding two secret digits the new average will become 235.25. Thus adjustment needs to add value 3 to R11 to equate the average value and bring it back to 236.

### 3.6 Authentication

At the receiving end the setgo-image passed through 2 x 2 mask labeled as $R_{00}$, $R_{01}$, $R_{10}$ and $R_{11}$. Mod operation on $R_{01}$, and $R_{10}$ by 4 to fetch the secret quaternary digits using the function given in (7).

$$\text{Secret digit1} = R_{01} \% 4$$
$$\text{Secret digit2} = R_{10} \% 4 \qquad (7)$$

After extraction of all the quaternary digits, the numbers are converted to radix 10 and the secret message/image is generated which when compared with the original secret image will authenticate the cover image.

## 4. RESULTS AND DISCUSSIONS

Ten PPM [10] images have been taken and applied on WASTIR to obtain results. All cover images are 512 x 512 in dimension and secret coin (k) used as authenticating image of 512 x 256 in dimension. The images are shown in figure 8.

Signal & Image Processing : An International Journal (SIPIJ) Vol.4, No.2, April 2013

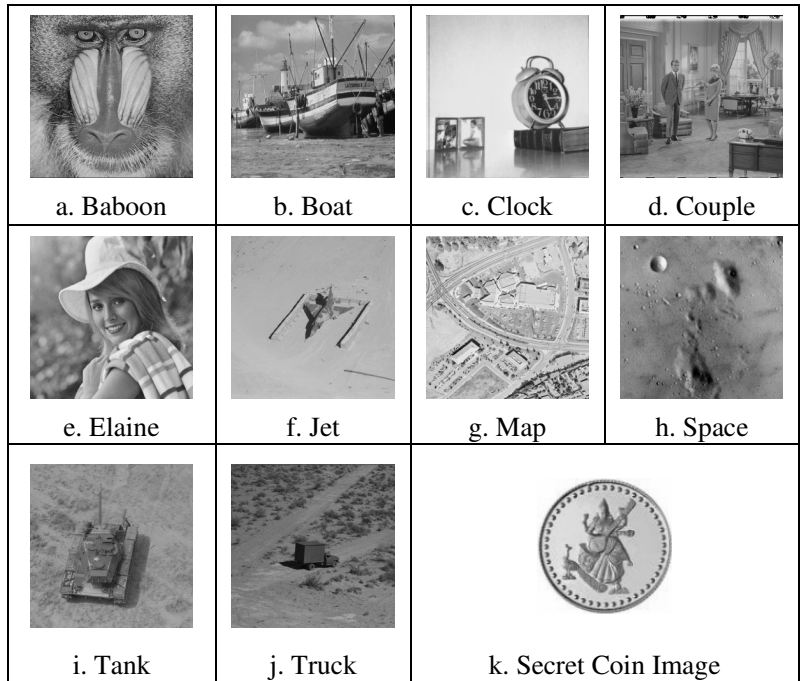

Figure 8.   Cover Images of dimension 512 x 512 and Secret Image  512 x 256

Table 1 shows the MSE, PSNR and IF for ten benchmark gray scaled images for embedding two quaternary digits into a 2 x 2 masks of translated image. As every digit requires maximum two bits to represent thus in total 1.0 bpB (bit per byte) of payload is supported by the proposed WASTIR algorithm. From table 1 it is clear that average MSE over ten images are 4.6478536, that of PSNR is 41.4588557 and IF is 0.999762.

Table I.     Statistical data on applying WASTIR over 10 gray scale translated images of 1024 x 1024 dimensions with secret coin as secret message/image.

| Cover Image 512 x 512 | MSE | PSNR | IF |
| --- | --- | --- | --- |
| (8.a) Baboon | 4.615559 | 41.488561 | 0.999752 |
| (8.b) Boat | 4.643509 | 41.462341 | 0.999756 |
| (8.c) Clock | 4.714600 | 41.396355 | 0.999875 |
| (8.d) Couple | 4.572992 | 41.528799 | 0.999727 |
| (8.e) Elaine | 4.631128 | 41.473935 | 0.999777 |
| (8.f) Jet | 4.796215 | 41.321817 | 0.999847 |
| (8.g) Map | 4.623446 | 41.481146 | 0.999865 |
| (8.h) Space | 4.697512 | 41.412125 | 0.999724 |
| (8.i) Tank | 4.506279 | 41.592623 | 0.999753 |
| (8.j) Truck | 4.677296 | 41.430855 | 0.999616 |

A comparative study has been made between WTSIC [3] and WASTIR in terms of Mean Square Error (MSE), Peak Signal to Noise Ratio (PSNR) and bits per Byte as in table 2. On increasing the bits per byte from 0.5 to 1.0, exactly double, the MSE increases by 0.599707 and PSNR increases by 0.61696 dB only which is acceptable compared to bits per Byte ratio.

Table II. Comparison between WTSIC, Horizontal sub-image method with proposed WASTIR.

| Image | WTSIC, Horizontal Orientation sub-image | | WASTIR | |
|---|---|---|---|---|
| | 0.5 (bit per Byte) | | 1.0 (bit per Byte) | |
| | MSE | PSNR | MSE | PSNR |
| Elaine | 3.786247 | 42.348714 | 4.631128 | 41.473935 |
| Boat | 3.696224 | 42.453221 | 4.643509 | 41.462341 |
| Clock | 4.426640 | 41.672001 | 4.714600 | 41.396355 |
| Map | 3.778519 | 42.357588 | 4.623446 | 41.481146 |
| Jet | 4.722733 | 41.388870 | 4.796215 | 41.321817 |

Comparison of the proposed technique has also been made between Yuancheng Li's Method [12], Region-Based watermarking methods [13] and ATFDD [14] as shown in table III. In all of the cases optimized results are obtained in proposed methods in terms of PSNR and payloads compare to existing techniques. But in case of comparison with WTSIC [3], on embedding into large images with increase of payload by more than eight times, the PSNR has decreased by only 0.5 dB.

Table III. Comparison between various techniques with proposed WASTIR

| Technique | Hiding Capacity (bytes) | Size of cover image | bpB (Bits per bytes) | PSNR (dB) |
|---|---|---|---|---|
| Li's Method[12] | 1089 | 257*257 | 0.13 | 28.68 |
| Region-Based[13] | 16384 | 512*512 | 0.5 | 40.79 |
| WTSIC[3] | 16384 | 512*512 | 0.5 | 42.04 |
| ATFDD[14] | 32768 | 512 * 512 | 1.0 | 36.70 |
| WASTIR | 131072 | 1024*1024 | 1.0 | 41.43 |

## 5. CONCLUSIONS

In this paper we have addressed the issue of hiding data in cover image in frequency domain, through discreet wavelet transformation technique (DWT). This method firstly resizes the image arbitrarily by up-sampling, followed by embedding. The original cover image is regenerates at the receiver by down sampling the image through transformation. The proposed technique provides a good fidelity in terms of PSNR, IF compared to existing techniques.

## ACKNOWLEDGEMENTS

The authors express deep sense of gratuity towards the Dept of CSE University of Kalyani where the computational resources are used for the work and the PURSE scheme of DST, Govt. of India.

Signal & Image Processing : An International Journal (SIPIJ) Vol.4, No.2, April 2013

## Authors

**Madhumita Sengupta**
M. Tech (CSE, University of Kalyani, 2010), pursuing Ph.D. as University Research Scholar (University of Kalyani), in the field of transform domain based image processing, steganography. Total number of publications is 14.

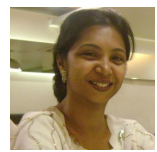

**Prof.  Jyotsna Kumar Mandal**
M. Tech.(Computer Science, University of Calcutta), Ph.D.(Engg., Jadavpur University) in the field of Data Compression and Error Correction Techniques, Professor in Computer Science and Engineering, University of Kalyani, India. Life Member of Computer Society of India since 1992 and life member of cryptology Research Society of India. Ex-Dean Faculty of Engineering, Technology & Management, working in the field of Network Security, Steganography, Remote Sensing & GIS Application, Image Processing. 26 years of teaching and research experiences. Nine Scholars awarded Ph.D. one submitted and eight are pursuing.  Total number of publications is more than two hundred seventy seven in addition of publication of five books from LAP Lambert, Germany.

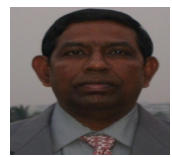